\documentclass{article}
\usepackage{amsmath}
\title{Alpha Blending with No Division Operations}
\author{Jerry R. Van\nobreak\hspace{.11em}Aken}
\usepackage{graphicx}
\usepackage{hyperref}
\begin{document}
  \maketitle

\begin{abstract}
\noindent Highly accurate alpha blending can be performed entirely with integer operations, and no divisions. To reduce the number of integer multiplications, multiple color components can be blended in parallel in the same 32-bit or 64-bit register. This tutorial explains how to avoid division operations when alpha blending with 32-bit RGBA pixels. An RGBA pixel contains four 8-bit components (red, green, blue, and alpha) whose values range from 0 to 255. Alpha blending requires multiplication of the color components by an alpha value, after which (for greatest accuracy) each of these products is divided by 255 and then rounded to the nearest integer. This tutorial presents an approximate alpha-blending formula that replaces the division operation with an integer shift and add---and also enables the number of multiplications to be reduced. When the same blending calculation is carried out to high precision using double-precision floating-point division operations, the results are found to exactly match those produced by this approximation. C++ code examples are included.
\end{abstract}

\section{Introduction}

The classic paper by Porter and Duff [1] describes how to do compositing of digital images that contain an alpha channel in addition to color channels. The per-pixel alpha components are idealized as real values that can vary over the normalized range 0 to 1, where 0 represents total transparency and 1 represents total opacity. This description is helpful for abstracting away the messy details of pixel arithmethic, but could lead the naive reader to assume that alpha blending requires division operations and might be best implemented using floating-point.

In practice, alpha values are commonly represented as compact integers. For example, a widely used format for digital images is RGBA32, for which each pixel is a 32-bit value that contains 8-bit red, green, blue, and alpha components, as shown in \mbox{Figure 1}.

\begin{figure}[h]
\centering
\includegraphics[width=10cm]{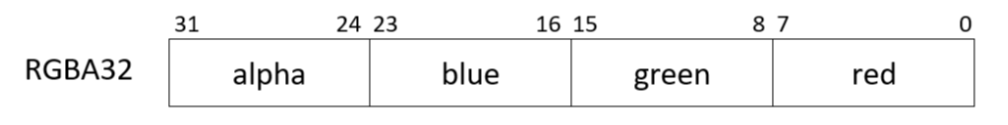}
\caption{RGBA32 pixel format}
\end{figure}

Each 8-bit component in Figure 1 is an unsigned integer in the range 0 to 255. To convert, for example, a raw 8-bit alpha component, $\alpha_{raw}$, to a normalized alpha value, $\alpha_{norm}$, where $0 \le \alpha_{norm} \le 1$, divide $\alpha_{raw}$ by 255. Thus, $\alpha_{raw} = 0$ corresponds to $\alpha_{norm}=0$, and $\alpha_{raw} = 255$ corresponds to $\alpha_{norm}=1$. The 8-bit red, green, and blue components can be normalized in similar fashion.

In terms of normalized components, the multiplication of, for example, red component $r_{norm}$ by alpha component $\alpha_{norm}$ to form product $r'_{norm}$ can be represented as
\begin{align*}
r'_{norm} &= \alpha_{norm} r_{norm} \\
               &= ( \frac{\alpha_{raw}}{255} ) ( \frac{r_{raw}}{255} )
\end{align*}
However, the result required from a typical alpha-blending operation is not the normalized value $r'_{norm}$, but rather the corresponding 8-bit component value, $r'_{raw}$, which might then be inserted into the red field of an RGBA32 pixel. The value $r'_{raw}$ can be expressed in terms of $r'_{norm}$ as
\begin{align}
r'_{raw} &= \lfloor 255\, r'_{norm} + \frac{1}{2}\, \rfloor \\
            &=  \lfloor \frac{\alpha_{raw} r_{raw}}{255} + \frac{1}{2}\, \rfloor \nonumber
\end{align}
where $\lfloor . \rfloor$ is the \emph{floor} function. Observe that the quantity $(\alpha_{raw} r_{raw}) / 255$ is rounded off to the nearest integer before truncation. If the division by 255 is carried out to the highest precision, the rounding error $-\frac{1}{2} \le \varepsilon \le \frac{1}{2}$ will be the sole source of error in $r'_{raw}$.

Of course, multiplications of 8-bit green and blue components by an 8-bit alpha value can be performed in similar fashion.

When optimizing for speed, programmers try to avoid unnecessary division operations. These operations are sequential in nature and typically require multiple clock cycles. To the aspiring graphics software designer, it may therefore appear that alpha blending requires a tradeoff between performance and accuracy. For example, if speed is prioritized over accuracy, the division by 255 might be replaced with division by 256, which can be implemented as a simple integer right-shift operation. On the other hand, at least one widely used open-source graphics library opts to sacrifice speed for accuracy by explicitly dividing by 255 [2].

However, this particular tradeoff between speed and accuracy is entirely unnecessary, as will be demonstrated in this tutorial. Consider that division of an 8-bit number by 255 (that is, by $2^8 - 1$) is a special case in which the quotient is known to be a repeating fraction in which the 8-bit number is repeated over and over again.

You can use your pocket calculator to verify that this special case holds for decimal numbers as well. Try dividing any two-digit decimal number by 99 (that is, by $10^2 - 1$). Here are some examples:
\begin{align*}
37 \div 99 &= 0.3737\,3737\,3737\,3737\,... \\
85 \div 99 &= 0.8585\,8585\,8585\,8585\,... \\
 9 \div 99 &= 0.0909\,0909\,0909\,0909\,...
\end{align*}
If your calculator supports a Programmer's Mode, you can verify that dividing any two-digit hexadecimal number by $\textrm{f\hspace{.07em}f}_{16}$ produces similar results. Some examples are
\begin{align*}
\textrm{4a} \div \textrm{f\hspace{.07em}f} &= \textrm{0.4a4a\,4a4a\,4a4a\,4a4a\,...} \\
\textrm{d7} \div \textrm{f\hspace{.07em}f} &= \textrm{0.d7d7\,d7d7\,d7d7\,d7d7\,...} \\
\textrm{ e} \div \textrm{f\hspace{.07em}f} &= \textrm{0.0e0e\,0e0e\,0e0e\,0e0e\,...}
\end{align*}
In other words, as soon as you are given an 8-bit number to divide by 255, you can immediately write down the resulting fraction to whatever precision you require. No division operation is required.

Thus, the division operation $(\alpha_{raw} r_{raw}) / 255$ in equation (1) can be converted to a multiplication of $r_{raw}$ by the repeating fraction $\alpha_{raw} / 255$. Equation (1) is then transformed to the following:
\begin{align}
r'_{raw} &=  \lfloor (\frac{\alpha_{raw}}{255}) r_{raw} + \frac{1}{2}\, \rfloor \\
             &=  \lfloor \big(\alpha_{raw} (2^{-8} + 2^{-16} + 2^{-24} + 2^{-32} + ...) \big) r_{raw} + \frac{1}{2}\, \rfloor \nonumber
\end{align}  

\section{Code examples}

To convert equation (2) into fast computer code that uses only integer operations, the repeating fraction $\alpha_{raw} / 255$ must be truncated so that the product of multiplication by 8-bit value $r_{raw}$ will fit in a processor register.

To keep things simple, we will truncate the repeating fraction to just 16 bits so that the fixed-point product $(\alpha_{raw}/255)r_{raw}$ can never exceed 24 bits, and will always fit in a 32-bit register. This approximation to equation (2) can be expressed as
\begin{align}
r'_{raw} &\approx \lfloor \big( \alpha_{raw} (2^{-8} + 2^{-16}) \big) r_{raw} + \frac{1}{2}\, \rfloor \\
             &= \lfloor 2^{-16} \Big( \big( \alpha_{raw} (2^8 + 1) \big) r_{raw} + 8000_{16} \Big) \rfloor \nonumber 
\end{align}
In this last expression, the constant $2^{-16}$ has been factored out of the quantity enclosed in the outer parentheses. Multiplication by this constant, which is implemented as a 16-bit right shift operation, is performed last so that the intermediate calculations are carried out with a full 24 bits of precision. The multiplication of $\alpha_{raw}$ by $(2^8 + 1)$ can be performed by shifting $\alpha_{raw}$ left by 8 bits and adding it back to its unshifted self.

The accuracy of the approximation in equation (3) can be tested by comparing the results it generates against those generated by using double-precision floating-point division operations to evaluate equation (1). If we run a test program that supplies all 65,536 possible combinations of $0 \le \alpha_{raw} \le 255$ and $0 \le r_{raw} \le 255$ as inputs, we will find that for all but 24 of these combinations, the values generated by this approximation exactly match those obtained by explicitly dividing by 255. For the 24 mismatches, the approximation always generates a value that is too low by just one. 

These encouraging test results suggest that a gentle nudge might be sufficient to increase the 24 mismatching values just enough that all 65,536 cases match exactly. In fact, the only change to equation (3) that’s needed to achieve 100-percent perfect matching is to increase the rounding constant from $8000_{16}$ to $8080_{16}$. With this small correction, the revised approximation is
\begin{align}
r'_{raw} &\approx  \lfloor 2^{-16} \Big( \big( \alpha_{raw} (2^8 + 1) \big) r_{raw} + 8080_{16} \Big) \rfloor 
\end{align}

In the following test program, which is written in C++, the \texttt{FastAlphaMult} function uses the approximation in equation (4) to calculate $r'_{raw}$. In contrast, the \texttt{SlowAlphaMult} function uses equation (1) to calculate $r'_{raw}$ to a high degree of precision, but does so at the cost of a double-precision floating-point division. The interested reader can compile and run this program to verify that the outputs from the two functions exactly match across all 65,536 possible combinations of the 8-bit \texttt{alpha} and \texttt{red} input values.

\begin{verbatim}
    #include <stdio.h>

    int FastAlphaMult(int alpha, int red)
    {
        alpha |= alpha << 8;
        red *= alpha;
        red += 0x8080U;
        return (red >> 16);
    }
    
    int SlowAlphaMult(int alpha, int red)
    {
        const double norm = 255.0;
        
        red = (alpha*red)/norm + 0.5;
        return red;
    }
    
    int test()
    {
        int misses = 0;
    
        for (int alpha = 0; alpha < 256; ++alpha)
        {
            for (int red = 0; red < 256; ++red)
            {
                int fastred = FastAlphaMult(alpha, red);
                int slowred = SlowAlphaMult(alpha, red);
    
                if (fastred != slowred)
                    ++misses;
            }
        }
        return misses;
    }
    
    void main()
    {
        printf("number of mismatches = %d\n", test());
    }
\end{verbatim}

The \texttt{FastAlphaMult} function above is a successful first attempt to evaluate equation (1) using only integer operations, and no divisions. However, alpha-blending operations typically require the blending of three or four components from each RGBA32 pixel. Performance would be further improved if multiple components could be alpha-blended in parallel in the same register.

Unfortunately, the calculation described in equation (4) requires 24 bits of register space per component, and so, with a register size of 32 bits, for example, only one component can be blended at a time. If this equation could be altered to require only 16 bits of register space per blended component, then two components (say, red and blue) could be blended in parallel in a 32-bit register, thereby reducing the number of multiplications per pixel from three (for red, green, and blue) or four (for example, if all four of the pixel’s components are multiplied by a stencil alpha value) to just two. Or, up to four components could be blended in parallel in a 64-bit register at the cost of a single multiplication.

To reduce the register space requirement from 24 to 16 bits, equation (4) can be transformed as follows:
\begin{align}
r'_{raw} &\approx  \lfloor 2^{-16} \Big( \big( \alpha_{raw} (2^8 + 1) \big) r_{raw} + 8080_{16} \Big) \rfloor \\
             &= \lfloor 2^{-16} \big( \alpha_{raw} r_{raw} (2^8 + 1) + 80_{16} (2^8 + 1) \big) \rfloor \nonumber \\
             &= \lfloor 2^{-16} \big( (\alpha_{raw} r_{raw} + 80_{16}) (2^8 + 1) \big) \rfloor \nonumber \\
             &= \lfloor 2^{-8} \big( (\alpha_{raw} r_{raw} + 80_{16}) (1 + 2^{-8}) \big) \rfloor \nonumber 
\end{align} 
If this last expression is to be evaluated using only integer operations, these operations should be performed in the following order. First, calculate the 16-bit sum of $\alpha_{raw} r_{raw}$ and rounding constant $80_{16}$. Second, to multiply this sum by $(1 + 2^{-8})$, shift the sum to the right by 8 bits and add it back to the unshifted sum. Finally, multiply this 16-bit intermediate result by $2^{-8}$ by shifting it right by 8 bits. The truncated final result is integer value $r'_{raw}$.

The following function, written in C++, implements the last expression in equation (5) in the manner just described. The return value is the 8-bit alpha-blended red component, $r'_{raw}$.  
\begin{verbatim}
    int FastAlphaMult2(int alpha, int red)
    {
        red *= alpha;
        red += 0x80U;
        red += red >> 8;
        return (red >> 8);
    }
\end{verbatim}
If this function, \texttt{FastAlphaMult2}, is inserted into the preceding test program in place of \texttt{FastAlphaMult}, it too will correctly match all 65,536 possible combinations of the 8-bit \texttt{alpha} and \texttt{red} input values.

The product \texttt{red*alpha} in the preceding \texttt{FastAlphaMult2} function never exceeds 16 bits. This fact suggests that it should now be possible to alpha-blend multiple 8-bit components in parallel in the same register.

The following function, \texttt{FastPremult}, demonstrates how to combine 8-bit red and blue components into a single 32-bit register and multiply them in parallel by the same 8-bit alpha value. This multiplication simultaneously applies equation (5) to both components. The \texttt{FastPremult} function, which is written in C++, premultiplies the color components in an array of RGBA32 pixels by their per-pixel alpha components.
\begin{verbatim}
    void FastPremult(UINT32 *pixel, int len)
    {
        for (int i = 0; i < len; ++i, ++pixel)
        {
            UINT32 color = *pixel;
            UINT32 alfa = color >> 24;
            UINT32 rb, ga;  
    
            color |= 0xff000000;
            rb = color & 0x00ff00ff;
            rb *= alfa;
            rb += 0x00800080;
            rb += (rb >> 8) & 0x00ff00ff;
            rb &= 0xff00ff00;
            ga = (color >> 8) & 0x00ff00ff;
            ga *= alfa;
            ga += 0x00800080;
            ga += (ga >> 8) & 0x00ff00ff;
            ga &= 0xff00ff00;
            *pixel = ga | (rb >> 8);
        }
    }
\end{verbatim}
In this function, the data type \texttt{UINT32} is defined to be an unsigned 32-bit integer. The first function parameter, \texttt{pixel}, is a pointer to the pixel array. The second parameter, \texttt{len}, is the array length. For each pixel in the array, the \texttt{for}-loop body first extracts the 8-bit alpha value from the pixel\footnote{For some applications, performance could be improved by checking for the special alpha values 255 and 0 before executing the rest of the loop body. However, this function, as it is written, enables numerical accuracy to be tested across \emph{all} values in the range 0 to 255.}. After the alpha value has been extracted, the alpha field in the \texttt{color} variable is set to 255 so that when this field is later multiplied by the previously extracted alpha, it will again equal this alpha.

Next, the red and blue components are isolated in variable \texttt{rb}, and \texttt{rb} is multiplied by the previously extracted alpha value. Then the constant \texttt{0x00800080}, which contains the rounding bits for the red and blue components, is added to the product in \texttt{rb} \emph{before} \texttt{rb} is shifted right by 8 bits and added to itself. The shifted value must be bitwise-ANDed with the mask \texttt{0x00ff00ff} before it is added back to \texttt{rb}; otherwise, the 8 LSBs of the shifted blue bits will clobber the 8 MSBs of the unshifted red bits.

The handling of the green and alpha components in variable \texttt{ga} is identical to that described for \texttt{rb}.

The performance of the \texttt{FastPremult} function above has been enhanced by reducing the number of multiplications to two, and by completely eliminating all divisions. Both enhancements are made possible by the use of equation (5). 

On some processors, the performance of the \texttt{FastPremult} function should improve if an optimizing compiler or human assembly programmer can load the mask constants into registers before the \texttt{for}-loop is entered.

For comparison, the \texttt{SlowPremult} function below performs the same premultiplication operation as the \texttt{FastPremult} function, but produces precise results by using double-precision floating-point division to implement equation (1). When these two functions are fed identical input arrays of RGBA32 pixels, their output values are identical.

\begin{verbatim}
    void SlowPremult(UINT32 *pixel, int len)
    {
        for (int i = 0; i < len; ++i, ++pixel)
        {   
            const double norm = 255.0;
            UINT32 color = *pixel;
            UINT32 red = color & 255;
            UINT32 grn = (color >> 8) & 255;
            UINT32 blu = (color >> 16) & 255;
            UINT32 alfa = color >> 24;
            
            red = (alfa*red)/norm + 0.5;
            grn = (alfa*grn)/norm + 0.5;
            blu = (alfa*blu)/norm + 0.5;
            *pixel = (alfa << 24) | (blu << 16) | (grn << 8) | red;
        }
    }
\end{verbatim}

\noindent Of course, the \texttt{SlowPremult} function doesn't have to be quite as slow as the listing above might imply.  The function could be sped up in several ways---for example, by using single-precision instead of double-precision floating-point, by calculating the factor \texttt{alfa/norm} before performing the multiplies, or by multiplying by the constant \texttt{1.0/255.0} instead of dividing by \texttt{255.0}. However, the \texttt{SlowPremult} function is written in this way to clearly show that it implements equation (1) without requiring the reader to perform mental gymnastics.

\section{Theory}

The interesting and useful properties of division by 255, as just discussed, derive from the following geometric series [3,\,4]:
\begin{align}
\frac{1}{1-x} = 1 + x + x^2 + x^3 + x^4 + ... \qquad \: \textrm{ for }x^2 < 1
\end{align}
To see how, consider the following division of a number $y$ by 255:
\begin{align*}
\frac{y}{255} = \frac{y}{256-1} = \frac{y/256}{1-1/256} = 2^{-8}y \big( \frac{1}{1-2^{-8}} \big)  
\end{align*}
The quantity in parentheses on the right can be expanded into a geometric series. To do so, simply substitute $2^{-8}$ for $x$ in equation (6), which yields
\begin{align}
\frac{y}{255} &= 2^{-8}y \big(1 + 2^{-8} + 2^{-16} + 2^{-24} + ...  \big)  \\
                     &= y ( 2^{-8} + 2^{-16} + 2^{-24} + 2^{-32} + ... ) \nonumber
\end{align}
For the special case of an integer $y$ restricted to 8 bits, this last expression reveals how a repeating fraction is formed as each successive multiplication of $y$ by the next higher multiple of $2^{-8}$ shifts the next instance of $y$ eight bits further to the right. However, this equation is equally valid for larger values of $y$.

\section{Some history}

The alpha-blending techniques presented in this tutorial are not new.

In 1995, Alvy Ray Smith [4] pointed out that division of a number $y$ by 255 can be expressed as multiplication of $y$ by a geometric series consisting of multiples of $2^{-8}$, as in equation (7) in the preceding section. Smith credits Jim Blinn with coming up with the best blending formula for 8-bit alpha and color components. Expressed as a C macro, Blinn's formula is
\begin{verbatim}
#define INT_MULT(a,b,t) ((t) = (a)*(b)+0x80, ((((t)>>8)+(t))>>8))
\end{verbatim}
where input parameters \texttt{a} and \texttt{b} are the alpha and color components, and parameter \texttt{t} is a variable that provides 16 bits of scratch storage for the intermediate calculation. The macro ``returns" (evaluates to) the 8-bit alpha-blended color component. Blinn's macro is equivalent to the \texttt{FastAlphaMult2} function presented in a previous section. Blinn presents his own derivation in [5].

Later, Blinn's \texttt{INT\_MULT} macro seems to have morphed into the following C macro, which is taken from the open-source PixMan code base [6] on the Apple website:
\begin{verbatim}
    #define FbByteMul(x, a) do {                         \
        CARD32 t = ((x & 0xff00ff) * a) + 0x800080;      \
        t = (t + ((t >> 8) & 0xff00ff)) >> 8;            \
        t &= 0xff00ff;                                   \
                                                         \
        x = (((x >> 8) & 0xff00ff) * a) + 0x800080;      \
        x = (x + ((x >> 8) & 0xff00ff));                 \
        x &= 0xff00ff00;                                 \
        x += t;                                          \
    } while (0)
\end{verbatim}
In this macro, input parameter \texttt{x} is an RGBA32 pixel value, input parameter \texttt{a} is an 8-bit alpha value, and the \texttt{CARD32} type is \texttt{unsigned int}. Note that \texttt{x} must be a variable (of at least 32 bits) because it appears on the left-hand side of the equals sign in the macro. The result, which is saved in \texttt{x}, is the alpha-blended RGBA32 pixel value.

The \texttt{FbByteMul} macro above is a close analog of the \texttt{FastPremult} function that was presented in a previous section, but performs a slightly different function\footnote{
The \texttt{FbByteMul} macro can be used in a couple of different ways. First, to premultiply the color components in pixel \texttt{x} by the pixel's alpha component, as was done inside the \texttt{for}-loop in the \texttt{FastPremult} function, (1) extract the pixel's alpha component into variable \texttt{a}, (2) bitwise-OR \texttt{x} with 0xff000000, and then (3) invoke the macro as \texttt{FbByteMul(x,a)}. Second, if pixel \texttt{x} is already in premultiplied-alpha format, and \texttt{a} is an 8-bit alpha value that is separate and distinct from the pixel's alpha component, then the invocation \texttt{FbByteMul(x,a)} multiplies the entire pixel, including its alpha component, by \texttt{a}. In this case, \texttt{a} could be an element in a \emph{matte} layer, as described in [1].
}. \texttt{FbByteMul} improves on Blinn's \texttt{INT\_MULT} macro by processing two color components in parallel, as was done in the \texttt{FastPremult} function.

The \texttt{FbByteMul} macro was originally part of the X server code base at X.Org, and only later migrated to PixMan. The macro was added to the X.Org Server repository in 2005 by Adam Jackson, with help from contributors Lars Knoll and Zack Rusin [7].

\section{References}

\begin{enumerate}
\item Porter, T., Duff, T. (July 1984). ``Compositing Digital Images." \emph{SIGGRAPH '84 Conference Proceedings}, \textbf{18}(3), 253-259.

\item See the \texttt{SDL\_Blit\_RGBA8888\_RGB888\_Blend} function in the script-generated \texttt{SDL\_blit\_auto.c} file on the SDL\;2 repository at \texttt{https://github.com/\hspace{.01em}libsdl-org/SDL/tree/120c76c84bbce4c1bfed4e9eb74e10678bd83120/\hspace{.01em}src/video}. As of January 2022 (commit \texttt{120c76c}), this function explicitly divides the integer product of an 8-bit alpha component and 8-bit color component by 255, but then loses a half-bit of accuracy by not rounding off this product before dividing. For this case (\emph{integer} division by 255), Blinn [5] shows that rounding should be done by adding 127, not 128.

\item Zwillinger, D. (Jan. 2018). ``Infinite Series." \emph{CRC Standard Mathematical Tables and Formulae}, Chapman \& Hall, 43.

\item Smith, A.R. (Aug. 1995). ``Image Compositing Fundamentals."  Technical Memo 4, Microsoft Corporation.

\item Blinn, J. F. (Nov. 1995). ``Three Wrongs Make a Right." \emph{IEEE Computer Graphics and Applications}, \textbf{15}(6), 90-93. 

\item See the \texttt{FbByteMul} macro in the \texttt{fbpict.h} file on the Apple source browser at \texttt{https://opensource.apple.com/source/WebCore/WebCore-3A109a/\hspace{.01em}platform/cairo/pixman/src/fbpict.h.auto.html}.

\item The \texttt{FbByteMul} macro was added to the \texttt{xserver/fb/fbpict.h} file in the X.Org Server repository in May 2005. The URL for the commit is \texttt{https://gitlab.freedesktop.org/xorg/xserver/-/commit/2de24db6\hspace{.01em}3eb65974ac547facf2a99aa4709d54b3}.
  
\end{enumerate}

\end{document}